\documentclass[journal=jacsat,manuscript=article]{achemso}
\usepackage{chemformula} % Formula subscripts using \ch{}
\usepackage[T1]{fontenc} % Use modern font encodings
\usepackage{amsmath,amssymb}

\author{Ruijun Wang}
\affiliation{State Key Laboratory of Optoelectronic Materials and Technologies, School of Electronics and Information Technology, Sun Yat-sen University, Guangzhou 510275, China}
\email{wangrj26@mail.sysu.edu.cn}
\alsoaffiliation{Institute for Quantum Electronics, ETH-Zurich, CH-8093 Zurich, Switzerland}
\author{Philipp Täschler}
\affiliation{Institute for Quantum Electronics, ETH-Zurich, CH-8093 Zurich, Switzerland}
\author{Zhixin Wang}
\affiliation{Institute for Quantum Electronics, ETH-Zurich, CH-8093 Zurich, Switzerland}
\author{Emilio Gini}
\affiliation{FIRST laboratory, ETH-Zürich, CH-8093 Zürich, Switzerland}
\author{Mattias Beck}
\affiliation{Institute for Quantum Electronics, ETH-Zurich, CH-8093 Zurich, Switzerland}
\author{Jérôme Faist}
\affiliation{Institute for Quantum Electronics, ETH-Zurich, CH-8093 Zurich, Switzerland}

\title[An \textsf{achemso} demo]
  {Monolithic integration of mid-infrared quantum cascade lasers and frequency combs with passive waveguides}
   
\abbreviations{IR,NMR,UV}

\begin{document}

\begin{abstract}
  Mid-infrared semiconductor lasers in photonic integrated circuits are of considerable interest for a variety of industrial, environmental and medical applications. However, photonic integration technologies in the mid-infrared lag far behind the near-infrared range. Here we present the monolithic integration of mid-infrared quantum cascade lasers with low-loss passive waveguides via butt-coupling. The passive waveguide losses are experimentally evaluated to be only 1.2±0.3 dB/cm with negligible butt-coupling losses. We demonstrate continuous-wave lasing at room temperature of these active-to-passive waveguide coupled devices. Moreover, we report frequency comb operation paving the way towards on-chip mid-infrared dual-comb sensors.
\end{abstract}

\section{Keywords}
quantum cascade laser, integrated photonics, frequency comb, mid-infrared,  spectroscopy, on-chip sensing

\section{Introduction}
The development of photonic integrated circuits (PICs) in the last decades has enabled high-speed optical transceivers and chip-scale photonic sensors \cite{Smit19,Tomb17}. On these platforms, the coupling of high-performance semiconductor lasers with low-loss passive waveguides is a vital step. Several approaches, including monolithic \cite{Cold00,Latk16,Liu18}, hybrid \cite{Guan18} and heterogeneous integration \cite{Wang19}, have been employed to integrate lasers on InP and silicon-based PICs. These PICs can offer compact, low-loss and high-quality factor laser feedback systems with a variety of filter components and enable novel solutions to realize advanced semiconductor laser sources. Such passive-active integration technologies for instance allow the design of widely tunable laser sources \cite{Cold00} and ultra-dense frequency combs \cite{Wang17}.

While advanced near-infrared PICs are readily available today, considerable efforts are devoted to bringing mid-infrared PICs to a similar degree of maturity. The mid-infrared wavelength range is of particular interest for spectroscopic sensing since it contains fingerprints of the most common molecular vibrations \cite{Gord17}. At present, quantum cascade lasers (QCLs) are the only semiconductor coherent sources that cover the entire mid-infrared spectrum in continuous-wave (CW) operation at room temperature \cite{Fais94,Yao12,Raze15}. In recent years, QCLs with multi-watt output powers in CW at room temperature have been realized \cite{Liu10,Bai10,Bism16}. Moreover, frequency comb operation was demonstrated\cite{Hugi12,Vill16,Hill19,Lu18,Kaza17,Meng20,Wang20}, enabling high resolution spectroscopy without moving parts\cite{Vill14}. 

Integrating QCLs with low-loss passive waveguides therefore provides a promising approach to realize chip-scale spectrometers for sensing applications. Correspondingly, the development of high-yield QCL integration platforms has been a long standing goal. Several strategies have been explored, including heterogeneous integration with silicon waveguides, hybrid integration in germanium PICs and monolithic integration with InGaAs waveguides on InP substrates \cite{Mont15,Spot16,Jung17,Nguy18,Go18,Rado19,Mali19,Jung19}. Integration on silicon can leverage on pilot-line fabrication processes for compact and low-cost photonic circuits. However, the low thermal conductivities of waveguide cladding materials and bonding interfaces strongly limit heat extraction from highly dissipative devices such as QCLs.  The first QCL integrated on silicon was demonstrated in 2016 and only lased in pulsed operation at room temperature \cite{Spot16}. These thermal issues can be avoided using hybrid integration on germanium \cite{Rado19}. However, in this case the optical mode mismatch between the QCL and germanium waveguide is a limiting factor for laser performance.

On InP-based monolithic integration platforms, on the other hand, both limitations can be overcome. Initially developed for optical communication applications, it now enables sophisticated PICs with more than 1000 components on a chip, including laser sources, photodetectors, modulators and various passive components \cite{Smit19}. Its extension towards longer wavelengths is particularly favorable as high-performance mid-infrared QCLs are routinely grown on InP substrates.  J. Montoya et al. demonstrated the monolithic integration of QCL active material with proton-implanted passive waveguides \cite{Mont15}. Due to passive waveguide losses around $\sim$7 dB/cm only pulsed operation could be achieved. Recently, lasing in pulsed operation was also reported by S. Jung et al. by homogeneously integrating 4.6 $\mu$m-wavelength QCLs with InGaAs passive waveguides via evanescent coupling \cite{Jung19}. While considerably lower waveguide losses around 2.2 dB/cm were measured,  the coupling efficiency between the active and passive waveguide sections was limited to $\sim$50\% in simulation. Moreover, the coupling efficiency was strongly dependent on the alignment accuracy in the lithography process and the dimensions of the waveguides.

In order to achieve continuous-wave lasing of monolithically integrated QCLs,  efficient optical coupling between active and passive waveguide sections and low passive waveguide losses are required. In this paper, we demonstrate such a system where light from an active QCL waveguide is coupled into a passive waveguide through a clean and smooth butt-joint. The waveguide core material is un-doped InGaAs with free-carrier absorption losses lower than 1 dB/cm. Since the active material and passive waveguide are grown in different steps, our approach offers high flexibility in epitaxial layer and waveguide structure. In particular, it can be applied to InP-based QCLs operating at any mid-infrared wavelength. In this work, we exemplify the approach based on QCLs lasing around 8 $\mu$m wavelength. 

The weak cavity dispersion observed in our devices enables frequency comb formation over a broad ($\sim$60cm$^{-1}$) optical spectrum. We experimentally verify comb operation by directly measuring the intermode coherence. 

\section{Design and Fabrication}
\begin{figure}[t]
\centering\includegraphics[width=12cm]{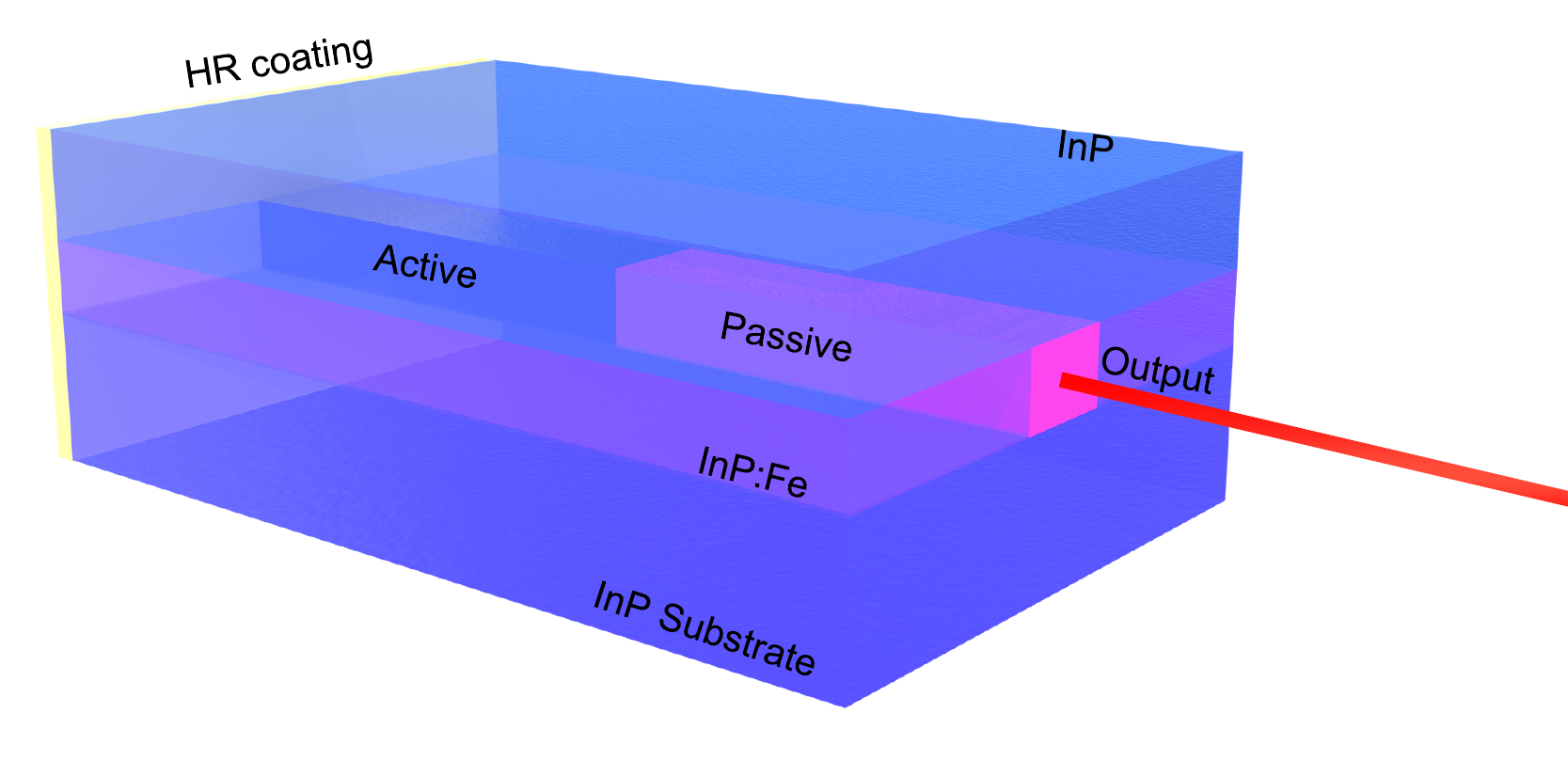}
\caption{Schematic of the monolithically integrated QCL. The light is coupled between the active and passive waveguide via a butt-joint.}
\label{Figure1}
\end{figure}
\begin{figure}[t]
\centering\includegraphics[width=10cm]{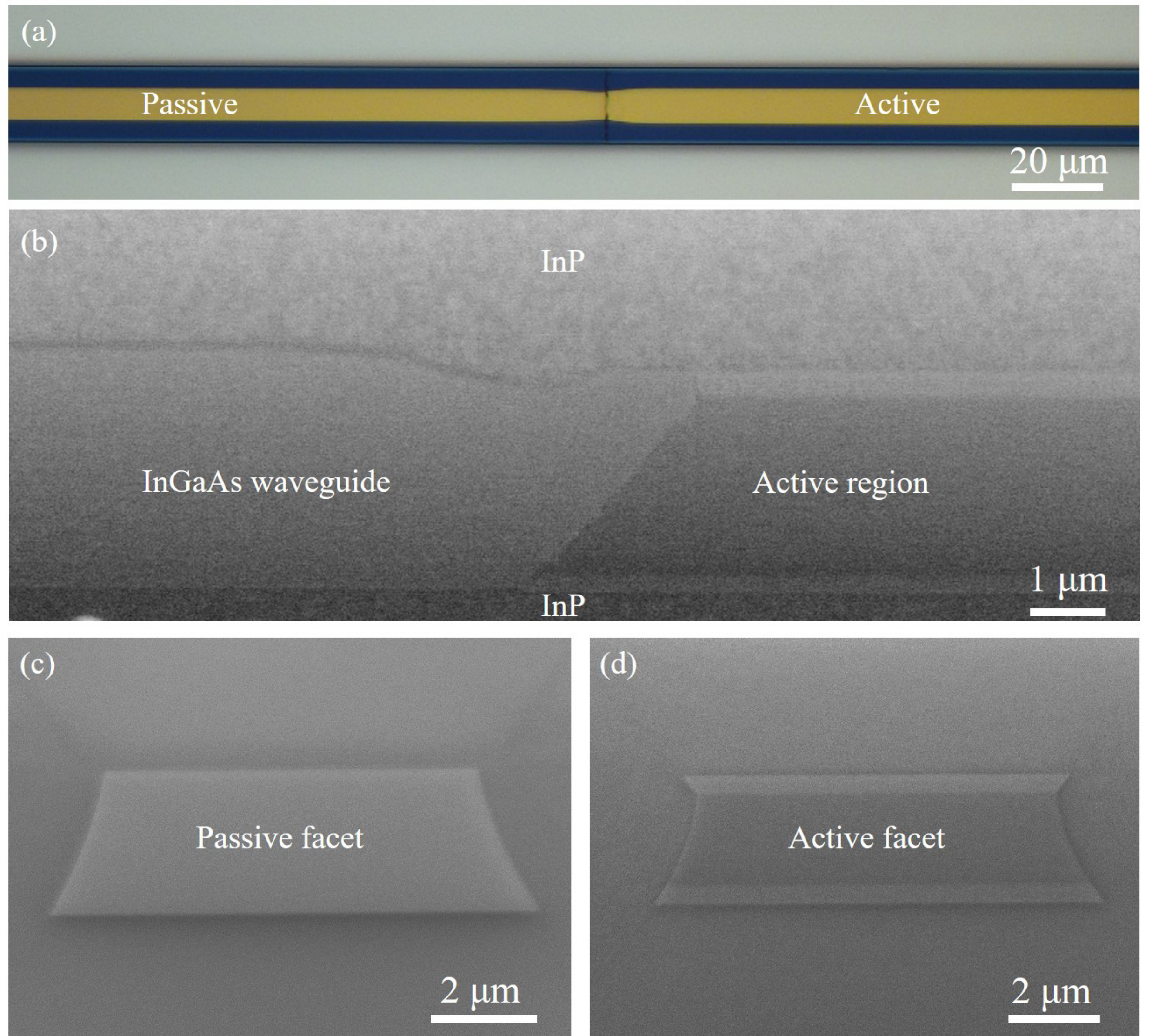}
\caption{(a) Top-view microscope image of the butt-coupled QCL waveguide after wet etching. (b) SEM image of the cross-section of the active-passive interface. (c)-(d) SEM images of the facet of the passive and active section. }
\label{Figure2}
\end{figure}
A schematic of our monolithically integrated QCL design is shown in Figure \ref{Figure1}. The InGaAs/AlInAs active region and InGaAs passive waveguide are cladded by InP materials to optically confine the light to the waveguide core. A layer of InP:Fe is grown on the sides of both waveguide sections for electrical insulation. The active and passive waveguide sections are vertically and horizontally aligned and exhibit only a small effective refractive index contrast at the interface which allows for a smooth waveguide transition.

The epitaxial layers constituting the active material are grown by molecular-beam epitaxy. The 2 $\mu$m thick active region is based on a strain-compensated InGaAs/AlInAs bound-to-continuum heterostructure which is embedded between two 300 nm thick layers of InGaAs. Straight stripes of active material with a width of 35 $\mu$m and a spacing of 400 $\mu$m are then defined by contact photolithography using a 400 nm thick SiO$_x$ layer as etching mask. Wet etching is performed using 1:1:3 H$_3$PO$_4$:H$_2$O$_2$:H$_2$O solution, which has a high selectivity of the QCL active region to InP. The anisotropic wet etching creates a 3.5 $\mu$m wide undercut below the SiO$_x$ hard mask layer.

Next, passive material (i.e. undoped InGaAs) of 2.3 $\mu$m thickness is grown on the wafer by metalorganic vapor phase epitaxy (MOVPE). The SiO$_x$ hard mask protects the active stripes from the epitaxial regrowth of InGaAs. A 50 nm thick InAlAs layer and 400 nm InP:Fe layer are then grown on the InGaAs layer to avoid currents flowing through the passive waveguide. After passive material regrowth, the hard mask layer is removed in hydrofluoric acid solution and a 500 nm thick n-doped InP cap layer (n= 5$\cdot$10$^{16}$ cm$^{-3}$) is regrown on the wafer. 

The homogeneously integrated QCLs are then processed using a standard buried heterostructure technique \cite{Sues16}. The active-passive waveguide ridges are defined by photolithography using a 400 nm thick SiO$_x$ layer as hard mask, followed by wet etching in 17:1:10 HBr:Br:H$_2$0 solution. Figure \ref{Figure2}a shows a processed waveguide with the passive and active sections labeled. A scanning electron microscope (SEM) image of the butt-joint cross section is depicted in Figure \ref{Figure2}(b), where we observe a smooth and clean waveguide transition between the two sections. The grown interface shows negligible thickness variations and material defects. 

A lateral layer of insulating InP:Fe is then selectively grown on the sides of the active and passive waveguides by MOVPE. The hard mask is removed and a 3 $\mu$m thick InP cladding layer is grown on the wafer again using MOVPE. In order to fully isolate the passive waveguide section from the electrical contact, a 200 nm thick SiN$_x$ layer is deposited. Afterwards, an ohmic contact on the top cladding layer is formed by Ti/Pt/Au deposition. The substrate is thinned down to 200 $\mu$m and Ni/Ge/Au is deposited as the backside contact on the InP substrate. In the end, the QCL wafer is cleaved into ridges with different passive-to-active ratios to form Fabry–Pérot (FP) lasers. Figure \ref{Figure2}(c) and (d) show SEM images of the active and passive facets, respectively. The devices are then mounted epitaxial-side down on AlN submounts on copper heat-sinks and are high-reflection (HR) coated (Al$_2$O$_3$/Au) on the rear facet belonging to the active waveguide section. In these devices, the laser cavity is formed between two cleaved facets. To integrate QCLs with PICs in the future, dry etching can be employed to fabricate laser feedback structures (e.g., Distributed Bragg Reflector) in the passive section after the monolithic integration process. In the fully integrated PICs, the light will be coupled from the QCL to the rest components by a partially reflective feedback structure or a direction coupler \cite{Smit19}.

\section{Monolithically integrated QCL}

\begin{figure}[t]
\centering\includegraphics[width=\linewidth]{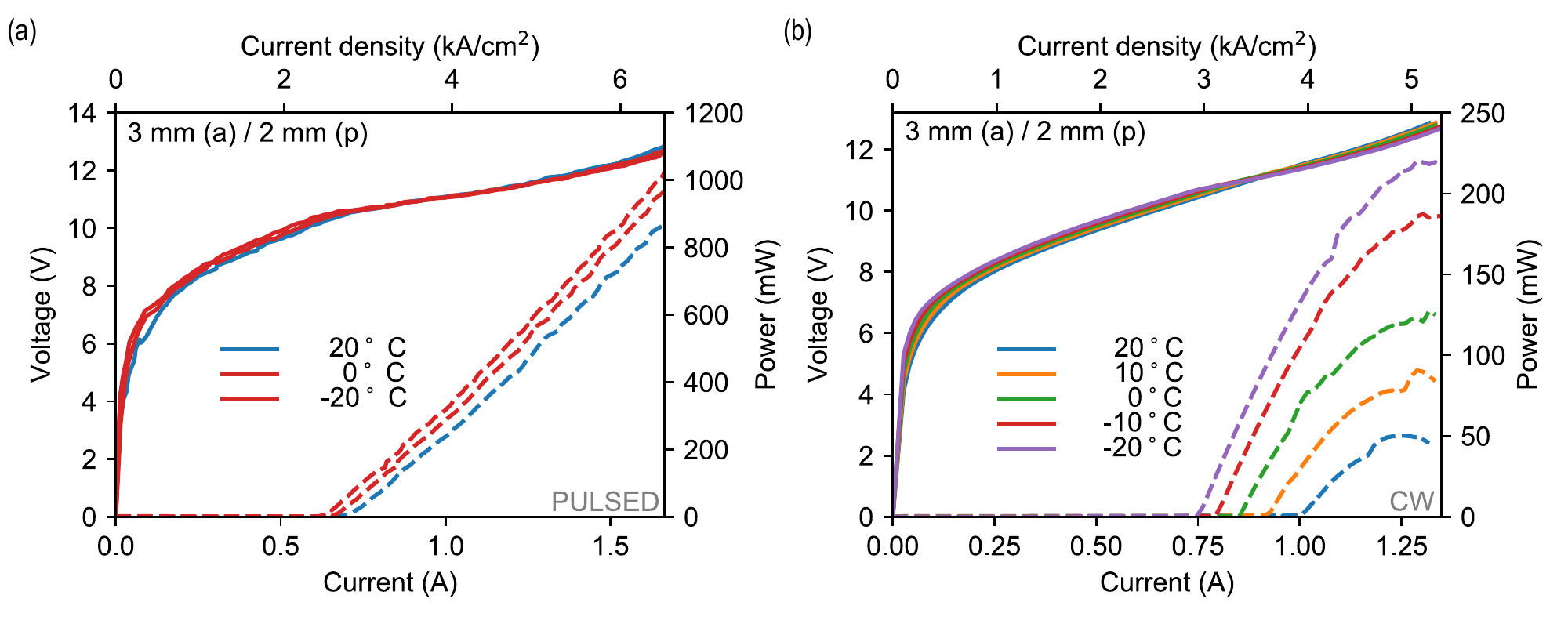}
\caption{(a) Pulsed LIV characteristic of a 3 mm active, 2 mm passive monolithically integrated QCL at different temperatures. (b) LIV characteristic of the same device as measured in CW operation. }
\label{Figure3}
\end{figure}

\begin{figure}[t]
\centering\includegraphics[width=0.8\linewidth]{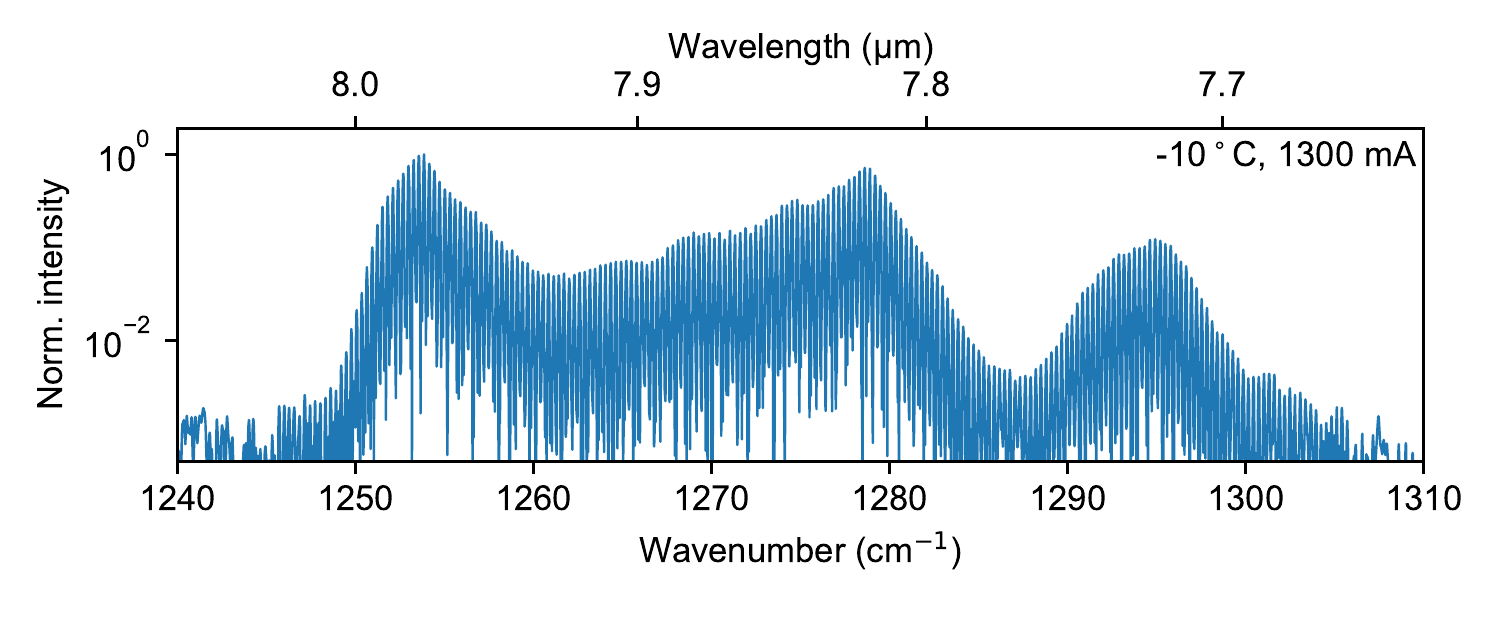}
\caption{Optical spectrum of the 3 mm active, 2 mm passive monolithically integrated device as obtained close to current rollover.}
\label{Figure4}
\end{figure}

The light-current-voltage (LIV) characteristics of a 3 mm active, 2 mm passive device in pulsed operation (100 ns pulse width, 100 kHz repetition rate) is shown in Figure \ref{Figure3} (a). At room temperature, we measure peak optical output powers up to 880 mW with a threshold current density of 2.72 kA/cm$^2$. Figure \ref{Figure3} (b) depicts the LIV curve as obtained in CW operation for the same device. Room temperature lasing with a maximum output power of 50 mW and a threshold current density of 3.93 kA/cm$^2$ is observed. LIV characteristics for devices with even longer passive sections (2.5 mm active and 4 mm passive, 1.5 mm active and 4.5 mm passive) are enclosed in the first section of Supplementary Materials .

The optical spectrum of the 3 mm active, 2 mm passive device close to current rollover at -10$^\circ$C is shown in Figure \ref{Figure4}. It was acquired using a Fourier transform infrared spectrometer with a spectral resolution of 0.075 cm$^{-1}$ and covers a bandwidth of more than 50 cm$^{-1}$ ($\sim$ 300 nm) in the 8 $\mu$m wavelength range. The measured mode spacing between adjacent spectral lines is evaluated to be $\sim$0.3 cm$^{-1}$ (8.7 GHz), which matches the free spectral range of the combined active-passive FP cavity of 5 mm length.

In the following, we determine the waveguide losses of our monolithically integrated QCLs by analyzing the slope efficiency of devices with different passive waveguide lengths. The overall waveguide loss can be divided into four separate terms: mirror losses $\alpha_M$, passive waveguide losses $\alpha_P$, active waveguide losses $\alpha_A$ and butt-coupling losses $\alpha_B$, which account for losses occurring at the active-passive interface \cite{Cold12}. The slope efficiency is directly related to these quantities via
\begin{equation}
    \eta_s = \frac{dP}{dI}=\eta_i\frac{N_p\hbar\omega}{e}\frac{\alpha_M}{\alpha_M+\alpha_A+\alpha_P+\alpha_B}
    \label{Equation1}
\end{equation}
where $e$ is the electron charge, $\eta_i$ the internal quantum efficiency, $N_p$ the number of periods in the QCL active region and $\hbar\omega$ the emitted photon energy. 

\begin{table}[t]
\centering
\begin{tabular}{cccc}
\multicolumn{2}{c}{\textbf{Waveguide (mm)}} & \multicolumn{2}{c}{\textbf{Slope efficiency (W/A)}} \\
Active          & Passive          & Before HR            & After HR            \\
\hline
2.5             & 0                & 1.00                 & 1.40                   \\
1.5             & 0.5              & 0.75                 & 1.14                   \\
3.5             & 0.5              & 0.68                 & 0.90                 \\
2.5             & 1                & 0.58                 & 0.86                 \\
3             & 2                  & 0.76                 & 1.02                    \\
1             & 2.5                & 0.43                 & 0.68               

\end{tabular}
\caption{Slope efficiencies of a single facet in pulsed operation at -20$^\circ$C for six devices with different active and passive waveguide lengths, as measured before and after HR coating.}
\label{Table1}
\end{table}

By considering pulsed light-current measurements of three integrated QCLs with different active and passive waveguide lengths, both before and after HR coating, Equation \ref{Equation1} can be used to determine the individual loss terms. The measured slope efficiencies of six devices before and after HR coating are reported in Table \ref{Table1}. Before the measurements of devices without and after HR coating, the power of a reference QCL device was measured to make sure the optical alignment of the setup is the same. It is found that the alignment accuracy is better than $5\%$ based on the measurement results of the reference device.

Before HR coating, mirror losses are given by $\alpha_M=-\ln{R^2}/2L$, where $R\approx27\%$ is the reflectivity at one facet as obtained from the calculated modal index of 3.2. Here, $L$ denotes the laser cavity length. After coating, the reflectivity of one facet approaches unity resulting in $\alpha_M\approx-\ln{R}/2L$. From these results and the observation that $\alpha_P$ and $\alpha_B$ vanish for the 2.5 mm only active device, $\alpha_A$ can be directly calculated. Applying Equation \ref{Equation1} to the rest devices leads to $\alpha_P=1.2\pm0.3$ dB/cm and $\alpha_B=0.05\pm0.02$ dB, respectively. Details about the calculation can be found in Supplementary Materials.

Numerical simulations predict a free-carrier absorption loss in the InGaAs passive waveguide of 0.7 dB/cm at 8 $\mu$m wavelength. This result indicates that the passive waveguide loss mainly originates from free-carrier absorption, with only minor contributions from scattering losses due to surface and sidewall roughness. It is worth noting that the QCL wafer used in this work was grown on an n-doped InP substrate (Si doped, $n=10^{17}cm^{-3}$). Further reduction in free-carrier loss could therefore be achieved by using a substrate with lower doping concentration. Compared with previous results in Ref. 31, the lower index contrast between the waveguide core (InGaAs, $n\approx3.4$) and cladding layer (InP, $n\approx3.1$) in the buried heterostructure configuration leads to the lower passive waveguide loss.

\section{Monolithically integrated QCL frequency comb}
Monolithic integration of mid-infrared frequency combs with passive waveguides offers the potential to realize fully integrated dual comb sensors. Besides, it provides new possibilities to design and control mid-infrared frequency combs. In the following, we evaluate our monolithically integrated QCL sources in terms of comb performance, with the example of a 3 mm active, 0.5 mm passive device.

\begin{figure}[t!]
\includegraphics[width=\linewidth]{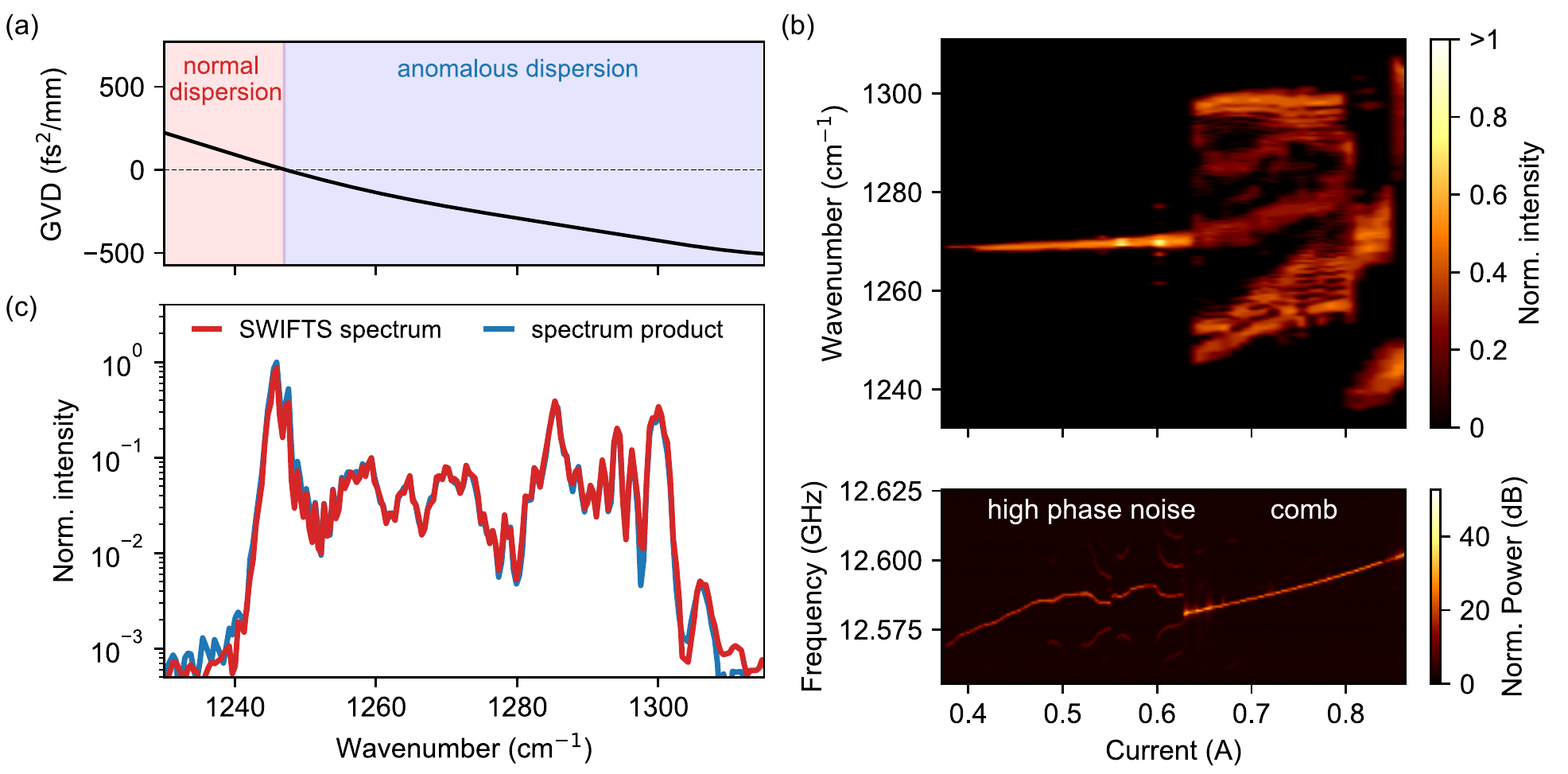}
\caption{(a) Second order dispersion as measured a few tens of milliamperes below lasing threshold. (b) Optical spectrum as a function of CW driving current for the monolithically integrated QCL. The lower subfigure displays the corresponding intermode beatnote which is measured in the microwave domain. (c) Spectral coherence as obtained by coherent beatnote interferometry.}
\label{Figure5}
\end{figure}

A prerequisite for comb formation is a low group velocity dispersion (GVD) over the full spectral bandwidth \cite{Vill16}. For the given device, the GVD was measured using a Fourier-transform-of-emission-spectrum approach \cite{Hofs99}. Measurements were carried out a few tens of milliamperes below lasing threshold and reveal a GVD of -450 fs$^2$/mm to 50 fs$^2$/mm within the lasing spectral region (Figures \ref{Figure5} (a)). Comb operation was previously demonstrated for QCL devices exhibiting similar values of GVD \cite{Wang20}. Measurements of GVD for devices with even longer passive section can be found in the Supplementary Materials Section.

Another indication for comb formation is a narrow ($\lesssim 1$ kHz) intermode beat \cite{Hugi12}, as it can be observed via the bias line of the device. In Figure \ref{Figure5} (b) we show the optical spectrum and corresponding beating signal, respectively. Below 620 mA and above 870 mA we identify a region of high-phase noise as represented by beattones of widths $\gtrsim$ 10 kHz. In between, a narrow sub-kHz beat indicates a phase-locked state and therefore the formation of a frequency comb. In this region, we observe spectral bandwidths of approximately 60 cm$^{-1}$.

In order to rigorously assess the coherence within the indicated region,  we performed coherent beatnote interferometry, also often referred to as SWIFTS \cite{Burg15,Sing18}. The high degree of mutual coherence among comb modes, is, up to a constant prefactor, shown in Figure \ref{Figure5} (c) by overlaying the SWIFTS spectrum with the spectrum product. During the measurement, the device was injection-locked to an external stabilized radio-frequency source.

\section{Conclusion}
In conclusion, we have presented the monolithic integration of mid-infrared QCLs with low-loss passive waveguides. The passive waveguide is fabricated by wet etching and a selective area growth technique. A clean and smooth butt-joint between the active and passive section with negligible optical losses is achieved. At the same time, waveguide losses of the passive segment are measured to be only 1.2±0.3 dB/cm. At room temperature, we report CW lasing of a 3 mm long active, 2 mm passive section device. Moreover, we demonstrated frequency comb operation of our monolithically integrated QCL sources. This work is a major advance towards mid-infrared PICs. In particular, our monolithic integration platform enables fully integrated on-chip spectroscopic sensors.

\subsection{Supporting Information}
This material is available free of charge via the internet at http://pubs.acs.org. LIV characteristics for devices with long passive sections (2.5 mm active and 4 mm passive, 1.5 mm active and 4.5 mm passive); calculation of the waveguide loss; measured GVD of monolithically integrated a QCL device with a 3 mm long active section and 2 mm long passive section.

\subsection{Funding Source}
The work was supported by the Swiss National Science Foundation under Grant Agreement No. 176584 (Combtrace).

\subsection{Acknowledgments}
The authors would like to thank Dr. Filippos Kapsalidis for helpful suggestions on processing.

\subsection{Notes}
The authors declare no conflicts of interest.

\end{document}